\documentclass[10pt]{article}
\usepackage{amsmath}
\usepackage{cite} 
\usepackage{indentfirst}
\usepackage{latexsym}
\usepackage{pstricks}

\long\def\ca#1\cb{} 

\newcommand{\becs}{\begin{cases}}
\newcommand{\bem}{\begin{matrix}}

\newcommand{\dya}[1]{|#1\rangle\langle#1|}

\newcommand{\encs}{\end{cases}}
\newcommand{\enm}{\end{matrix}}

\newcommand{\ket}[1]{|#1\rangle }
\newcommand{\lbrk}[1]{\left\{\vrule height #1cm depth #1cm width 0pt\right.}

\newcommand{\ot}{\otimes }

\newcommand{\st}{\sqrt{2}}

\newcommand{\Tr}{{\rm Tr}}

\newcommand{\vb}{\,|\,}

\newcommand{\AC}{{\mathcal A}}
\newcommand{\BC}{{\mathcal B}}

\newcommand{\HC}{{\mathcal H}}

\newcommand{\al}{\alpha }
\newcommand{\bt}{\beta }

\newcommand{\lm}{\lambda }

\def\Ac{A }
\def\Bc{B }
\def\ac{a }
\def\bc{b }

\begin{document}

\title{Nonexistence of Quantum Nonlocality}

\author{Robert B. Griffiths
\thanks{Electronic mail: rgrif@cmu.edu}\\ 
Department of Physics,
Carnegie-Mellon University,\\
Pittsburgh, PA 15213, USA}
\date{Version of 16 April 2013}
\maketitle  

\begin{abstract}
  What violations of Bell inequalities teach us is that the world is quantum
  mechanical, i.e., nonclassical.  Assertions that they imply the world is
  nonlocal arise from ignoring differences between quantum
  and classical physics. 
\end{abstract}

In a vast and ever growing literature, of which
\cite{BrGs11,Nrsn11,CrHW11,XiWL12,VrBr12,Plzl12,BrGs12,Bnao12,Vrao13,Cvs13,
  Gsn13} is but a selection of relatively recent examples, the claim is made
that quantum mechanics is \emph{nonlocal} because its predictions, confirmed by
experiment, violate various Bell inequalities.  Entire books
\cite{dEsp06,Mdln11b} have been written in support of this thesis, and one is
often given the impression that nonlocality is not only part of current quantum
mechanics, but will be a part of any future theory that gives the same
predictions for certain experimental situations.  Here we shall argue that, on
the contrary, derivations of Bell inequalities are based on assumptions or
intuitions that are appropriate in \emph{classical}, not quantum physics, and
that when a fully consistent quantum mechanical approach is employed to
describe the phenomena of interest, evidence for the supposed nonlocality of the
quantum world evaporates.  Quantum mechanics is consistent with special
relativity, and there is not the slightest indication of any superluminal
action at a distance. By contrast, derivations of Bell inequalities are based
on intuitive ideas more appropriate to the classical than to the quantum world,
and it is the failure to recognize this that has led to mistaken claims for
nonlocality.
In the following discussion we focus on statistical correlations for bipartite
quantum systems, as it is in this arena that the main discussion of (supposed)
quantum nonlocality has occurred.  However, the basic idea can be extended to
correlations in multipartite quantum systems, for which there have been similar
claims of nonlocality, e.g. \cite{TnYO13}, in a fairly obvious way.

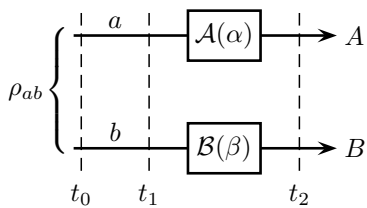
\begin{figure}[h]
$$
\begin{pspicture}(-1,-1)(4,2) 
\newpsobject{showgrid}{psgrid}{subgriddiv=1,griddots=10,gridlabels=6pt}
\def\lwd{0.035} 
\psset{
labelsep=2.0,
arrowsize=0.150 1,linewidth=\lwd}
\def\dput(#1)#2#3{\rput(#1){#2}\rput(#1){#3}}
\def\rectc(#1,#2){%
\psframe[fillcolor=white,fillstyle=solid](-#1,-#2)(#1,#2)}
\def\hdg{0.35} \def\squ{%
\psframe[fillcolor=white,fillstyle=solid](-\hdg,-\hdg)(\hdg,\hdg)}
\def\lwdsh{0.020}
\def\tnline(#1,#2,#3)#4{
\rput(#1,#2){\psline[linestyle=dashed,linewidth=\lwdsh](0.0,-0.3)(0.0,#3)}
\rput(0,-0.7){\rput[B](#1,#2){#4}}}
	%
\psline{->}(0.0,0.0)(3.5,0.0)
\psline{->}(0.0,1.5)(3.5,1.5)
\dput(2.0,1.5){\rectc(.5,.35)}{$\AC(\al)$}
\dput(2.0,0){\rectc(.5,.35)}{$\BC(\bt)$}
\rput(-0.2,0.75){$\lbrk{0.8}$}
\tnline(0.1,0,1.8){$t_0$}
\tnline(1.0,0,1.8){$t_1$}
\tnline(3.0,0,1.8){$t_2$}
\rput[l](3.6,1.5){$\Ac$}
\rput[l](3.6,0){$\Bc$}
\rput[r](-0.4,0.75){$\rho_{\ac\bc}$}
\rput[b](0.55,1.6){$\ac$}
\rput[b](0.55,0.1){$\bc$}
\end{pspicture}
$$
\caption{Particles $\ac$ and $\bc$ prepared in a state $\rho_{\ac\bc}$ at time
  $t_0$ are later, between $t_1$ and $t_2$, measured by devices $\AC$ and
  $\BC$, measurement settings $\al$ and $\bt$, resulting in
  outcomes $\Ac$ and $\Bc$.}
\label{fgr1}
\end{figure}

Consider a situation, shown schematically in Fig.~\ref{fgr1}, in which two
particles $\ac$ and $\bc$, Hilbert spaces $\HC_\ac$ and $\HC_\bc$, are prepared
at some location and then move away from each other to be measured at a later
time by devices $\AC$ and $\BC$, with respective measurement outcomes $\Ac$ and
$\Bc$.  Following the initial preparation there is no interaction between any
of the four systems involved, the two particles and the two measuring devices,
except for the measurements themselves, during which $\AC$ interacts with $\ac$
and $\BC$ with $\bc$.  The measurement devices $\AC$ and $\BC$ are prepared
with measurement settings $\al$ and $\bt$, respectively, which determine the
sorts of measurements they carry out. For example, if $a$ is a spin-half
particle, the $\al=z$ setting for $\AC$ would result in a measurement of the
$z$ component $S_{az}$ of its spin, with macroscopic outcomes $A=\pm 1$
corresponding to $S_{az}=\pm 1/2$ in units of $\hbar$.  The \emph{quantum
  formula} for the conditional probabilities of outcomes $\Ac$ and $\Bc$, given
measurement settings $\al$ and $\bt$, is
\begin{equation}
 \Pr(\Ac,\Bc\vb \al,\bt) = 
\Tr\bigl\{[R(\Ac|\al)\ot S(\Bc|\bt)] \rho_{\ac\bc}\bigr\}.
\label{eqn1}
\end{equation}
Here $\rho_{\ac\bc}$ is the density operator on the Hilbert space
$\HC_\ac\ot\HC_\bc$ of the two particles, as determined by unitary time
evolution from the time they were prepared up until just before the
measurement. For a fixed $\al$ the $R(\Ac|\al)$ for different values $A$ form a
POVM, or in the simplest case a (projective) decomposition of the identity for
particle $a$; similarly the $S(\Bc|\bt)$ are POVMs for particle $b$. Thus as
well as being positive operators they sum to their respective identities,
\begin{equation}
 \sum_\Ac R(\Ac|\al) = I_\ac,\quad \sum_\Bc S(\Bc|\bt) = I_\bc.
\label{eqn2}
\end{equation}
We are using a notation in which $\Pr()$ is the probability of whatever is
indicated by the symbols inside the parentheses, conditioned on whatever is to
the right of the vertical bar.  In general the probabilities given by
\eqref{eqn1} are \emph{correlated} in the sense $\Pr(\Ac,\Bc\vb\al,\bt)\neq
\Pr(\Ac\vb\al,\bt)\Pr(\Bc\vb\al,\bt)$. This is to be expected since the
particles were close together at an earlier time when they could have
interacted with each other, or they were prepared by some localized
apparatus. Correlations arise in classical as well as quantum models, and in
and of themselves do not indicate any sort of nonlocality.

Rather than \eqref{eqn1}, derivations of Bell inequalities employ what we call
the \emph{Bell formula}
\begin{equation}
 \Pr(\Ac,\Bc\vb\al,\bt) = 
\sum_\lm \Pr(\lm) \Pr(\Ac\vb \al,\lm)\Pr(\Bc\vb\bt,\lm),
\label{eqn3}
\end{equation}
or something essentially equivalent to it.  Here $\lm$ refers to a
state-of-affairs that existed at an appropriate point in time, say $t_1$ in
Fig.~\ref{fgr1}, after the particles were prepared and before they were
measured; see the discussions in \cite{Bll90c,Nrsn11}.  Often $\lm$ is
referred to as a hidden variable, but a more precise identification is not
needed; simply assuming that probabilities have the form \eqref{eqn3} leads to
certain results, in particular Bell inequalities, not always satisfied by
probabilities given by the quantum formula \eqref{eqn1}.  By now there are many
experimental measurements confirming the validity of \eqref{eqn1} in situations
where it can be proved that these probabilities do not satisfy \eqref{eqn3}.
The claim we wish to dispute is that because probabilities given by the quantum
formula \eqref{eqn1} are not always of the Bell form \eqref{eqn3}, therefore
quantum mechanics, or the world it describes, is nonlocal.

What additional assumptions beyond those of standard quantum mechanics used in
writing down \eqref{eqn1} are needed to arrive at (conditional) probabilities
that can be written in the form \eqref{eqn3}?  We know no general answer, but
it is helpful to examine two particular cases.
The first is that in which
$\rho_{\ac\bc}$ is a \emph{separable} density operator, one that can be
written in the form
\begin{equation}
 \rho_{\ac\bc} = \sum_\lm \Pr(\lm) \bar\rho_\ac(\lm)\ot \bar\rho_\bc(\lm),
\label{eqn4}
\end{equation}
where for each $\lm$, $\bar\rho_\ac(\lm)$ and $\bar\rho_\bc(\lm)$ are density
operators for particles $\ac$ and $\bc$, and $\Pr(\lm)$ a normalized
probability distribution.  Inserting \eqref{eqn4} in \eqref{eqn1} yields
\begin{equation} \Pr(\Ac,\Bc\vb\al,\bt) = \sum_\lm \Pr(\lm)\,
  \Tr_\ac\bigl[R(\Ac|\al)\bar\rho_\ac(\lm)\bigr]\,
  \Tr_\bc\bigl[S(\Bc|\bt)\bar\rho_\bc(\lm)\bigr],
\label{eqn5}
\end{equation}
which is obviously of the form \eqref{eqn3}.  

A second route from \eqref{eqn1} to \eqref{eqn3}, one that does not require
$\rho_{\ac\bc}$ to be separable, is to assume that for all $\Ac$, $\Ac'$,
$\al$, and $\al'$ it is the case that
\begin{equation}
 [R(\Ac|\al), R(\Ac'|\al')] = 0,
\label{eqn6}
\end{equation}
which is to say all the POVM operators for measurements of particle $a$ commute
with each other.  Then \eqref{eqn1} leads to \eqref{eqn5} (and thence to
\eqref{eqn3}) by the following construction. It follows from \eqref{eqn6} that
there is an orthonormal basis $\{\ket{\lm}\}$ of $\HC_\ac$, consisting of kets
we label by $\lm$, such that in this particular basis all matrices representing
$R(\Ac|\al)$ (for all $\Ac$ \emph{and} for all $\al$) are diagonal. Using this
basis when evaluating the trace in \eqref{eqn1} allows one to write the right
side in the form \eqref{eqn5} with
\begin{equation}
\Pr(\lm) = \Tr\bigl\{( [\lm]\ot I_\bc)\rho_{\ac\bc}\bigr\},\quad
\bar\rho_\ac(\lm) = [\lm], \quad 
\bar\rho_\bc(\lm) =\Tr_\ac\bigl\{([\lm]\ot I_\bc)\rho_{\ac\bc}\bigr\}/\Pr(\lm),
\label{eqn7}
\end{equation}
where $[\lm] = \dya{\lm}$ is the projector corresponding to $\ket{\lm}$.
Of course, if the $S(\Bc|\bt)$ all commute among themselves the analogous
construction again leads to \eqref{eqn5} if one uses an appropriate orthonormal basis
$\{\ket{\lm}\}$ of $\HC_\bc$. 

Both of these routes for obtaining \eqref{eqn3} from \eqref{eqn1} have a strong
``classical odor.'' A separable density operator \eqref{eqn4} represents, more
or less by definition, a state that is not entangled, and it is generally
agreed that entanglement is a feature of quantum mechanics not anticipated in
classical physics.  Similarly, the assumption in \eqref{eqn6} that certain
operators commute with each other suggests a classical point of view: one way
to locate the boundary between classical and quantum physics is to put it at
the point where one can no longer assume that physical properties are
represented by quantities that commute with each other.

But there are also situations where conditional probabilities cannot be
expressed in the form \eqref{eqn3}.  Probably the best known is the one in
which $\rho_{ab}$ is the projector onto the entangled singlet state
\begin{equation}
 \ket{\psi_0} = (1/\st)(\ket{0,1} - \ket{1,0})
\label{eqn8}
\end{equation}
of two spin-half particles.  Here $\ket{0,1}$ is an eigenstate of $S_{az}$ with
eigenvalue $+1/2$ in units of $\hbar$, and an eigenstate of $S_{bz}$ with
eigenvalue $-1/2$.  As is well known, there are certain choices of POVMs that
satisfy neither \eqref{eqn6} nor its counterpart for the $S(\Bc|\bt)$, for
which the collection of conditional probabilities $\Pr(\Ac,\Bc\vb\al,\bt)$
given by \eqref{eqn1} \emph{cannot} be written in the form \eqref{eqn3} because
they violate a Bell inequality derived from \eqref{eqn3}.  (See, e.g., the
discussion in Sec.~VI C of \cite{Nrsn11}.)  Is it the case that
some peculiar quantum nonlocality manifests itself in this situation, but is
somehow invisible in situations in which \eqref{eqn4} or \eqref{eqn6} apply?
Perhaps, but that seems unlikely.

The derivation of \eqref{eqn3} given by Bell in
\cite{Bll90c}, or in the extremely careful analysis by Norsen \cite{Nrsn11},
is based on a certain intuition about the
property or properties labeled by $\lm$ at time $t_1$,
Fig.~\ref{fgr1}.
%
Reasoning about the state of affairs that exists in a system between the time
of its preparation and the time it is measured is more subtle in quantum than
in classical mechanics, and carelessness can easily lead to various paradoxes,
such as those discussed in Chs.~20 to 25 of \cite{Grff02c}.  In particular,
there is often more than one quantum sample space, which is to say more than
one (projective) decomposition of the quantum Hilbert space identity operator,
which may be of physical interest.  For example, if the singlet state
\eqref{eqn8} is prepared and the time development of the particle state between
$t_0$ and $t_1$ is trivial (no interaction between the particles, no magnetic
fields), textbook quantum mechanics tells us to use $\ket{\psi_0}$ (i.e., the
projector onto this ray) at time $t_1$, and this is what is used by Bell and
Norsen, see the quotation following Eq.~(14) in \cite{Nrsn11}, as if there were
no other choice.  But that is not the only possibility.  Assume that Alice is a
competent experimentalist and has set up an apparatus $\AC$ to measure
$S_{az}$.  Then it is rather natural for her to ascribe to this component of
the spin angular momentum of particle $\ac$ at the earlier time $t_1$ the
property revealed by her later measurement output at time $t_2$, and this is
appropriate if she employs the proper quantum analysis; see the discussion of
measurements in Chs.~17 and 18 of \cite{Grff02c}.  This is the type of
reasoning made all the time in experimental particle physics when earlier
properties of a particle are inferred from observing it in some detector, and
it is fully compatible with a quantum description of microscopic properties in
terms of Hilbert space projectors, not some sort of hidden variables.

But a quantum description in which $S_{az}$ is $+1/2$ or $-1/2$ is incompatible
with one in which the joint state of the two particles is the singlet state
\eqref{eqn8}, at least if the latter is thought of as a property rather than a
pre-probability (in the notation of Sec.~9.4 of \cite{Grff02c}): the projectors
do not commute.  The problems raised by noncommuting projectors representing
different quantum properties have no classical analog, and thus classical
intuition cannot be relied on to resolve them; one has to analyze them using
quantum mechanics in a consistent way.  This problem often arises in situations
involving quantum measurements, and Bell, as is evident from \cite{Bll90},
lacked a satisfactory approach for dealing with it.

Even if Bell's (and Norsen's) attempts to derive \eqref{eqn3} are flawed
through reliance on classical intuition in a situation where quantum mechanics
is needed, this still leaves open the possibility that quantum theory gives
rise to some peculiar nonlocality.  To rule out this possibility one needs
positive and fully consistent \emph{quantum} arguments to the effect that in a
situation such as in Fig.~\ref{fgr1}, the choice of measurement, say $\al$ for
$\AC$, has no influence on the properties of the distant particle $\bc$.  Such
arguments exist.  The case of the singlet state \eqref{eqn8} was explored in
 Ch.~23 of \cite{Grff02c}, and again in \cite{Grff11b}.  In \cite{Grff11}
it was shown that quantum theory implies a principle of
\emph{Einstein locality}:
\begin{quote}
Objective properties of isolated individual systems do not change when 
something is done to another non-interacting system.
\end{quote}
Thus the \emph{absence of interaction} assumed when writing down the quantum
formula \eqref{eqn1} is the key to genuine quantum locality.  By contrast, the
Bell formula \eqref{eqn3} should be seen not as a criterion for locality, but
as identifying certain limited cases in which classical intuition retains some
of its validity in a quantum context.

In summary, the quantum world is local, quantum mechanics agrees with special
relativity, and violations of Bell inequalities do not indicate any kind of
nonlocality.  Instead they point to certain features of the quantum world, the
only real world known to us at the present time, which are inconsistent with
classical physics, and thus with various intuitions derived from classical
physics.  There is, of course, no harm in investigating the circumstances under
which quantum correlations do or do not have a particular form; this is a
mathematical question with the potential of giving rise to valuable physical
insights.  However, this potential seems most likely to be realized if the
misleading idea of nonlocality is replaced by a serious effort to understand
the implications of noncommutativity of the projectors representing quantum
properties.

I thank Dr. T. Norsen for correspondence regarding these matters.  The research
reported here received financial support from the National Science Foundation
through Grant PHY-1068331.

\end{document}